\documentclass[twocolumn,showpacs,superscriptaddress,amsmath,amssymb]{revtex4}
\usepackage{graphicx}
\usepackage{dcolumn}
\usepackage{bm}
 
\def\bd{
\begin{document}} \def\ed{\end{document}}
\def\bmp{\begin{minipage}} \def\emp{\end{minipage}}
\def\bcc{\begin{center}} \def\ecc{\end{center}}     \def\npg{\newpage}
\def\underbar#1{$\setbox0=\hbox{#1} \dp0=1.5pt \mathsurround=0pt
   \underline{\box0}$}   \def\ub{\underbar}    \def\ul{\underline}
\def\dit{\item{-}} \def\i{\item} \def\us{^{(s)}}
\def\btm{\begin{itemize}} \def\etm{\end{itemize}}
\def\btbl{\begin{tabular}} \def\etbl{\end{tabular}}
\def\btbb{\begin{tabbing}} \def\etbb{\end{tabbing}}
\def\bea{\begin{eqnarray}} \def\eea{\end{eqnarray}}
\def\beq{\begin{equation}} \def\eeq{\end{equation}}
\def\bbb{\begin{thebibliography}{9} \itemsep=-1mm}
\def\ebb{\end{thebibliography}} \def\bb{\bibitem}
\def\bpic{\begin{picture}(260,240)} \def\epic{\end{picture}}
\def\akgt{\noindent{Acknowledgements\ \ }}
\def\hph{\hphantom}  \def\r#1{$^{[#1]}$} \def\fc{\frac}  \def\cf{{\it cf.}\;}
\def\nd{\noindent}   \def\pa{\parindent}  \def\ie{{\it i.e.}\;}
\def\hs{\hskip} \def\vs{\vskip} \def\hf{\hfill} \def\eg{{\it e.g.}\;}
\def\cl{\centerline} \def\ob{\obeylines}  \def\ls{\leftskip}
\def\bs{\boldsymbol}
\def\f{\left} \def\g{\right} \def\e{{\rm e}} \def\d{{\rm d}}
\def\vf{\varphi} \def\pl{\partial} \def\cov{{\rm cov}} \def\ch{{\rm ch}}
\def\la{\langle} \def\ra{\rangle} \def\EE{e$^+$e$^-$\;}
\def\dt{\delta} \def\ta{\theta} \def\dt{\delta} \def\Dt{\Delta}
\def\vf{\varphi}  \def\ve{\varepsilon} \def\gm{\gamma}  \def\Gm{\Gamma}
\def\gmy{\gamma_y} \def\gmpt{\gamma_{\pt}}  \def\gmf{\gamma_\vf}
\def\r{\rho}
\def\pt{{p_{\rm t}}}  \def\yct{y_{\rm cut}} \def\kt{k_{\rm t}}
\def\ktct{{\kt}_{\rm cut}}   \def\nbr{\nonumber}
\def\ifmath#1{\relax\ifmmode #1\else $#1$\fi}%
\def\rc{\ifmath{{\mathrm{c}}}}  \def\cut{\ifmath{{\mathrm{cut}}}}
\def\rF{\ifmath{{\mathrm{F}}}}  \def\rK{\ifmath{{\mathrm{K}}}}
\def\rp{\ifmath{{\mathrm{p}}}}  \def\rt{\ifmath{{\mathrm{t}}}}
\def\LAB{\ifmath{{\mathrm{LAB}}}}  \def\cut{\ifmath{{\mathrm{cut}}}}
\newcommand{\cinst}[2]{$^{\mathrm{#1}}$~#2\par}
\newcommand{\crefi}[1]{$^{\mathrm{#1}}$}
\newcommand{\crefii}[2]{$^{\mathrm{#1,#2}}$}
\newcommand{\crefiii}[3]{$^{\mathrm{#1,#2,#3}}$}
\newcommand{\HRule}{\rule{0.5\linewidth}{0.5mm}}
\def\js{Jetset\,7.4\ }   \def\hw{Herwig\,5.9\ }
\def\ktctprod{\ktct^{\mbox{\footnotesize jet-prod}}}
\def\ktctdev{\ktct^{\mbox{\footnotesize jet-dev}}}

\bd

\title{A Monte Carlo study on the production scale \\
and internal structure of jets in high energy collisions}

\author{Chen Gang\footnote{E-mail: chengang1@cug.edu.cn}}
\affiliation{ Department of Physics, China University of
Geosciences, Wuhan 430074}
\author{Yu Meiling, Liu Lianshou\footnote{E-mail: liuls@iopp.ccnu.edu.cn}}
\affiliation{Institute of Particle Physics, Huazhong Normal
University, Wuhan 430079}


\begin{abstract}
The production scale and internal structure of jets produced in
high energy collisions are studied using \js and \hw Monte Carlo
generators. Two scales are found. One is the {\it jet-development
scale}, which determines the size of the jet developed from a
mother-parton. The other one is the {\it jet-production scale},
the jets produced with this scale are the most consistent with QCD
jet-production dynamics and will provide the most reliable
dynamical information about their mother-partons.
\end{abstract}


\maketitle

\section{Introduction}

The basic theory of strong interaction ------ Quantum
Chromo-Dynamics (QCD) possesses the special property of both
asymptotic freedom and color confinement. For this reason the
partons (quarks and/or gluons) produced in high energy collisions
have to be turned into hadrons before they can be observed in
experiments. When the virtuality $Q$ of a parton is high enough, the
produced hadrons will preserve the momentum of the mother-parton,
representing themselves as a cone around the moving direction of
mother-parton and is referred to as {\it jet of hadrons}, or simply
{\it jet}. Jets, being experimentally observable, are widely used as
a tool for the experimental investigation of the physical properties
of partons and their interaction dynamics.

In 1975 a two-jet structure was observed in \EE annihilation
experiments at $\sqrt s \leq 6$ GeV~\cite{Hanson}. This has been
taken as the experimental confirmation of the production of a pair
of quark-antiquark, moving back to back in \EE collisions, as
predicted by  the parton model~\cite{JELIS}.

As energy increases the quark or anti-quark can emit a hard, \ie
high transverse momentum, gluon producing a third jet. This
astonishing prediction of QCD was confirmed by experiments, when a
third jet was observed in \EE collisions at $\sqrt s = 17$ -- 30
GeV~\cite{expobservethirdjet}. This observation has been
recognized as the first experimental evidence of gluon.

The situation in hadron-hadron or nucleus-nucleus collisions are
somewhat more complicated due to the existence of a large
background. However, since the basic interaction ------ QCD is the
same, the partons produced in these collisions, if having high
enough transverse momenta, will also produce jets. The production of
jets in hadron-hadron collisions was widely studied in the 80 --
90$^{\rm th}$ of the last century~\cite{UA1, UA2} and has been used
as an effective way for extracting the strong coupling constant
$\alpha_{\rm _S}$~\cite{alfas}.

In last century the nucleus-nucleus (heavy ion) collisions were
performed in SPS at CERN. The corresponding experiments were fixed
target ones with $\sqrt{s_{NN}}$ being lower than 20 GeV. No jet
production is expected to be observable at these energies. Only
starting from this century, when the first heavy ion collider RHIC
at BNL successfully run Au-Au collision at $\sqrt{s_{NN}}$= 200
GeV, jet production becomes available. Jet physics is an important
part of RHIC program. The observation of energy
lose~\cite{observeElose} of hard jets passing through the medium
produced in the collision is one of the main achievements of RHIC,
which is referred to as jet quenching, and is taken as one of the
signals for the formation of a hot dense matter in relativistic
heavy ion collisions~\cite{jetqsignal}. Furthermore, jet is
recognized as a powerful tool for studying the properties of the
produced new form of matter~\cite{WangWang}.

In view of the highly importance of jet physics, it is necessary to
study the definition and structure of jet in more detail. These are
the aim of the present paper.

\section{The scales in jet definition}

The definition of jet depends on scale.

Theoretically~\cite{theorydefofjet} jet is defined as a certain
fraction $\ve$ of energy deposited in a cone with opening angle
$\dt$ around some axis. Here $\ve$ and $\dt$ determine the scale
of jet.

Experimentally, jets can be identified through some jet-finding
process, \eg the Jade~\cite{Jade} or Durham~\cite{Durham}
algorithm. In these processes there is a parameter $\yct$, which
in case of the Durham algorithm is related to the cut in relative
transverse momentum $\kt$~\cite{Dokshitzer}\;\cite{natural}
\beq 
\ktct = \sqrt{\yct}\cdot \sqrt{s},\eeq where $s$ is the c.m.
energy squared. The relative transverse momentum $\kt$ between two
particles $i$ and $j$ is defined as~\cite{Dokshitzer} \beq
{\kt}_{ij}=2 \min (E_i,E_j)\;\sin\f(\frac{\ta_{ij}}{2}\g). \eeq
When a particle has a $\kt$ relative to an existing jet smaller
than $\ktct$ then it is grouped into this jet. The value of
$\ktct$ is the scale of jet.

Alternatively, jets can also be identified by the cone
algorithm~\cite{Coneargorithm}. Fixing some direction as jet axis,
the {\it pseudo-rapidity} $\eta$ and {\it azimuthal angle} $\vf$
plane are constructed, where $\eta$ is defined as
$\eta=-\fc{1}{2}\ln \tan \ta$, $\ta$ is the angle with jet axis. A
parameter $R=\sqrt{\eta^2+\vf^2}$ is defined for each particle.
Particles with $R\leq R_0$ is identified as a jet, where $R_0$ is
the scale of jet.

In relativistic heavy ion experiments, people usually use a high
transverse momentum particle as trigger to define a
jet~\cite{triggerjet}. The threshold of the trigger momentum is the
scale of jet.

In all the above cases it is generally believed that how to choose
the scale for jet production is a matter of definition or a matter
of taste. The {\it jet production scale} is considered to be
arbitrary in a large extent. Different scales give rise to
different jets and you can choose one within a wide range that
fits your requirement.

In the present paper this problem will be revisited. We will try
to answer the question: in which sense can we consider the scale
of jet as arbitrary, and whether or not there is a {\it most
reasonable} scale of jet that is consistent with the physics of
jet production and development in QCD.

\section{The production and development of jet}

\begin{figure}
\includegraphics[width=1.8in]{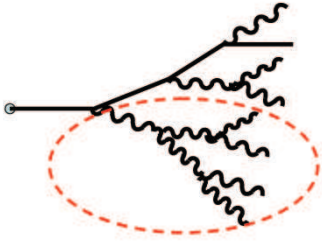}
\caption{\label{Fig. 1} A schematic sketch of the development of
parton shower from a mother quark or anti-quark.}
\end{figure}

For simplicity, let us take \EE collision as example.

According to QCD the quark (anti-quark) produced in high energy
\EE collision will emit gluon, and the emitted gluon will in turn
emit further gluons or, with lower probability, convert into
quark-antiquark pairs. In this way, a parton shower is produced,
turning finally into hadronic jet. The development of a parton
shower from a mother-quark, or anti-quark, is sketched
schematically in Fig.\;1. (The other parton shower produced by the
accompanied anti-quark, or quark, in the collision is not shown in
the figure.) The development of parton shower is above a scale
$Q\geq Q_0\sim 1$\;GeV, which is the scale discriminating
perturbative and non-perturbative QCD. When the virtuality $Q$ is
smaller than $Q_0$, perturbative calculation becomes unapplicable
and the partons hadronize into final state hadrons
non-perturbatively.

It can be seen from the figure that the first emitted gluon will
develop into a sub-parton-shower, surrounded in the figure by a
dashed ellipse. When the transverse momentum of the gluon is low
this sub-parton-shower will be mixed with the other partons and
represent itself as a part of a unique jet. However, when the
transverse momentum of the gluon is high enough, the produced
sub-parton-shower might be separated from the main part, forming
by itself a new jet. The transverse momentum cut that determines
whether the gluon can be considered as having produced a separate
jet or as being melted in the unique jet is the {\it scale for the
production of jet}.


Is there any value which is the most reasonable  scale for the
production of jet according to the jet production and development
processes in QCD?

To answer this question consider the development of a jet from a
mother-parton through parton shower. According to the symmetry of
QCD Lagrangian, the parton shower has only one privileged
direction, \ie the direction of the mother-parton momentum.
Therefore, dynamically a single jet should possess an axial
symmetry.

Then let us consider the dynamical symmetry of the q$\bar {\rm
q}$g system, \ie the quark anti-quark together with the first
emitted gluon, produced in \EE collision. This is the basis of a
3-jet event. According to the symmetry of QCD Lagrangian, the
gluon emission is isotropic, so the dynamical symmetry of 3-jet
events is spherical.

Therefore, the most reasonable scale for jet-production is
characterized by the dynamical symmetry of a single jet being
axial, while that of the 3-jet events being spherical.

In a 2-jet event there are two jets flying back to back with a
unique axis. Therefore, the dynamical symmetry of 2-jet events is
the same as that of a single jet, \ie axial.

However, in order to check the dynamical symmetry of a system, the
final state particle distributions could not be used. The symmetry
properties of the final state particle distributions are
controlled not only by the dynamics of elementary processes but
also by the process of hadronization and kinematics. They are
strongly influenced by the latter, which will enshroud the
dynamical symmetry of the system. In order to study the dynamical
symmetry of a system, getting rid of the influence of
hadronization and kinematics, the symmetry of {\it dynamical
fluctuations} should be used instead of that of the final state
{\it particle distributions}.

The jets from 2-jet events possessing axial symmetry with respect
to dynamical fluctuations have been studied in
Ref.\;\cite{CGLlsPRD} and have been given the name ------ {\it
circular} jet. The scale of circular jet is found to be around
$\ktct^{\rm circ} = 6.32\pm 0.03$\;GeV for the 2-jet sample from
\js while $4.28\pm . 0.02$\;GeV for that from \hw.

In order to check the dynamical symmetry of 3-jet events, two
event samples each with 2,000,000 \EE events are generated from
\js and \hw at $\sqrt s=$91.2 GeV, respectively. In Fig's.\;2 (a)
and (b) are shown the variation with $\yct$ ($\ktct$) of the three
parameters of dynamical fluctuations
------ $\gmy, \gmpt, \gmf$ (\cf
Appendix) in the 3-jet events from the above-mentioned two
samples. It can be seen that at $\ktct^{\mbox{\footnotesize
3-jet}}=6.03\pm 0.06$\;GeV for \js and $3.98\pm 0.03$\;GeV for \hw
the three $\gm$'s go together and the system possesses spherical
symmetry.

\begin{figure}
\includegraphics[width=2.8in]{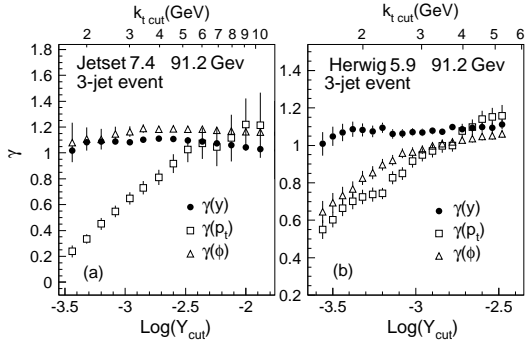}
\caption{\label{Fig. 2}  The variation of the parameter $\gamma$
with $\yct$ ($\ktct$) in 3-jet events from (a) \js and (b) \hw at
$\sqrt s=$91.2 GeV;}
\end{figure}

Thus we see that the jet produced at the scale $4\sim 6$\;GeV
possesses the dynamical symmetry expected by QCD. This is the
appropriate or {\it most reasonable} scale of jet production that
is consistent with the dynamics of jet production and development
in QCD.

\section{The longitudinal and transverse distributions of particles
inside jets}

Let us now turn to discuss the particle distributions inside jet
obtained from the Durham jet-algorithm with various $\ktct$'s. For
this purpose 2- and 3-jet events are selected from the two event
samples from \js and \hw mentioned above for six values of
$\ktct$: $\ktct=2,4,6,8,10,12$ GeV. Then one jet is taken out from
each event and the internal structure of this jet is studied.

Firstly, the momenta of all the particles in the jet are summed up
to $\bs p_{\rm jet}$, defined as the {\it jet momentum}. The
direction of $\bs p_{\rm jet}$ is defined as the {\it
longitudinal} direction and the directions perpendicular to it are
the {\it transverse} directions. The rapidity $y$ and transverse
momentum $\pt$ are defined along these directions as usual.

In Fig.\;3 are shown the rapidity distributions inside a single
jet in 2-jet events corresponding to six $\ktct$ values. It can be
seen that the rapidity distribution increases sharply from $y=0$
to 1, then turns to a slower rise until a maximum pick around
$y=2$--4 is reached. The pick moves leftward as the increasing of
$\ktct$. The rapidity distributions inside a single jet in 3-jet
events, shown in Fig.\;4, have a similar behavior.

\begin{figure}
\includegraphics[width=2.8in]{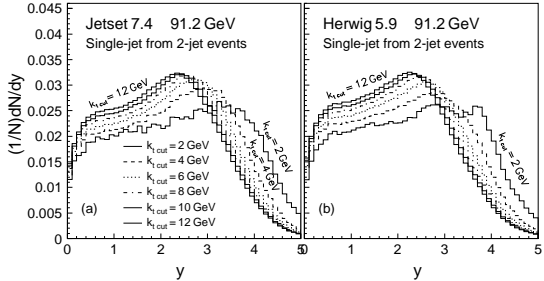}   
\caption{\label{Fig. 3} The rapidity distribution inside a single
jet in 2-jet events corresponding to six $\ktct$ values, from (a)
\js and (b) \hw at $\sqrt s=$91.2 GeV. }
\end{figure}

\begin{figure}
\includegraphics[width=2.8in]{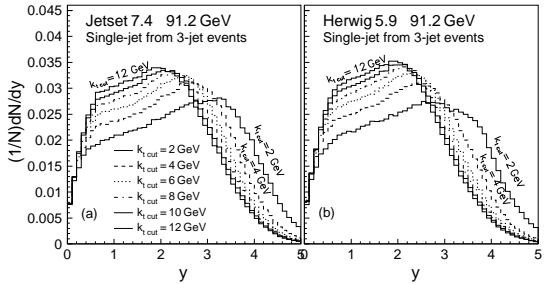}   
\caption{\label{Fig. 4} The rapidity distribution inside a single
jet in 3-jet events corresponding to six $\ktct$ values, from (a)
\js and (b) \hw at $\sqrt s=$91.2 GeV. }
\end{figure}

The particles with rapidity at or to the right of the pick are the
leading particles of the corresponding jets. Their momenta along
the jet axis, taking as $z$ axis, are about $p_z\geq 10$ GeV/$c$.
This can be used as the scale for triggering jets.

Contrary to the dependence of rapidity distribution on $\ktct$,
the relative transverse momentum $\kt$ distribution inside a
single jet turns out to be insensitive to $\ktct$, \cf Fig.'s\; 5
and 6. Here we use the Jade definition for $\kt$~\cite{Jade} \beq
{\kt}_{ij}^{\rm jade}=
2\sqrt{E_iE_j}\;\sin\f(\fc{\ta_{ij}}{2}\g),\eeq which is basically
the same as Durham $\kt$ but is symmetric with respect to $i$ and
$j$. This definition is more natural when we study the internal
structure of jet, instead of absorbing a particle into the main
part of jet in jet-algorithm.

The insensitivity of $\kt$ distribution on $\ktct$ means that
there is a scaling property in the transverse direction inside
jet. Note that the value of $\ktct$ is the parameter, which
determines how to group particles to jets. A particle that has a
$\kt$ with respect to an existing jet less than $\ktct$ belong to
that jet, while those with $\kt$ larger than $\ktct$ belong to
another jet. Thus $\ktct$ is the lower limit of the distance
between two jets and at the same time the upper limit of the size
of jet. Therefore, it is natural to expect that the size of jet
will increase with the increasing of $\ktct$. The transverse
scaling shown in Fig's.\;5, 6 is contrary to this expectation.

\begin{figure}
\includegraphics[width=2.8in]{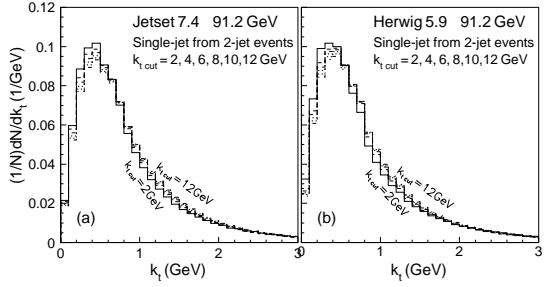}   
\caption{\label{Fig. 5} The relative transverse momentum  $\kt$
distribution inside a single jet in 2-jet events, corresponding to
six $\ktct$ values, from (a) \js and (b) \hw at $\sqrt s=$91.2
GeV. }
\end{figure}
\begin{figure}
\includegraphics[width=2.8in]{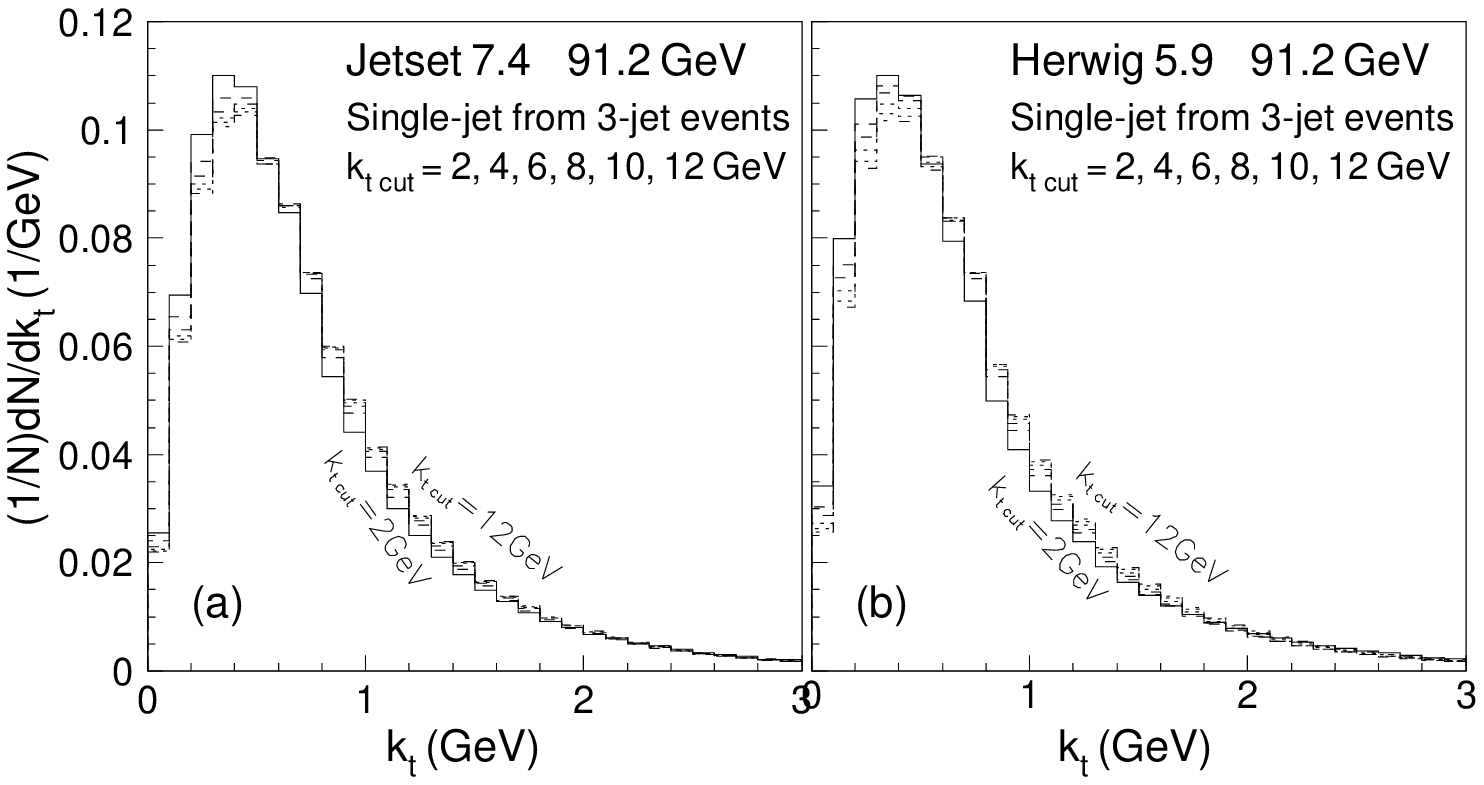}   
\caption{\label{Fig. 6} The relative  transverse momentum $\kt$
distribution inside a single jet in 3-jet events, corresponding to
six $\ktct$ values, from (a) \js and (b) \hw at $\sqrt s=$91.2
GeV. }
\end{figure}
\begin{figure}
\includegraphics[width=2.8in]{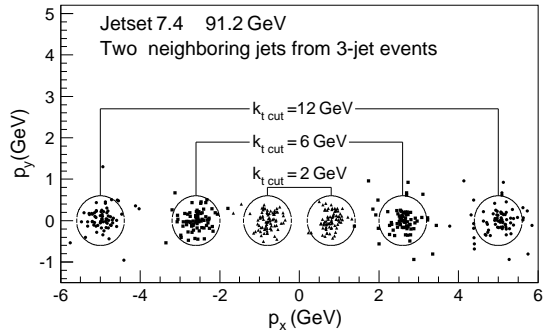}   
\caption{\label{Fig. 7} The scattering plots of the particles in
two neighboring jets of 3-jet events, corresponding to three
$\ktct$ values. }
\end{figure}

\def\pta{{p_{\rm t}a}}    \def\ptb{{p_{\rm t}b}}
In order to be more intuitive, we show in Fig.\;7 the scattering
plot of the particles in two neighboring jets of 3-jet events,
corresponding to three $\ktct$ values. Two jets $a$ and $b$ are
{\it neighboring} means that the distance $\sqrt{
({\pt}_{ax}-{\pt}_{bx})^2+({\pt}_{ay}-{\pt}_{by})^2 }$ of their
axes being the smallest among all the jet-distances in the event.
A reference frame is constructed, using the line ${\pt}_a$
${\pt}_b$ as $x$ axis and putting the origin at the middle point
of the line segment $\overline{{\pt}_a {\pt}_b}$. In order to
increase statistics 10 events with two neighboring jets separated
for the same distance are selected and the results from these
events are superposed in Fig.\;7. It can clearly be seen from the
figure that the size of jet is universal, independent of the
distance between jets.

\begin{figure}
\includegraphics[width=2.8in]{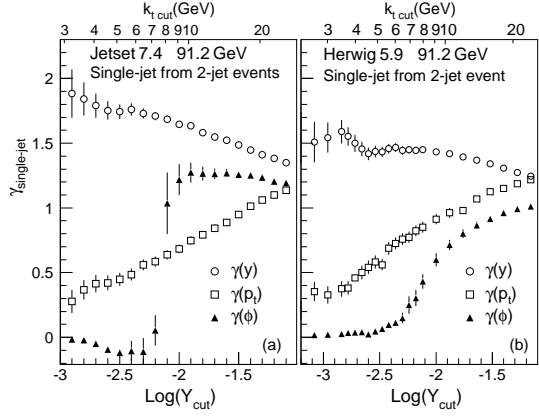}   
\caption{\label{Fig. 8}  The variation of the parameters
$\gamma(y)$, $\gamma(\pt)$, $\gamma(\vf)$  with $\yct$ ($\ktct$)
in a single jet of 2-jet events  from (a) \js and (b) \hw at
$\sqrt s=$91.2 GeV, calculated in jet-axis frame.}
\end{figure}
\begin{figure}
\includegraphics[width=2.8in]{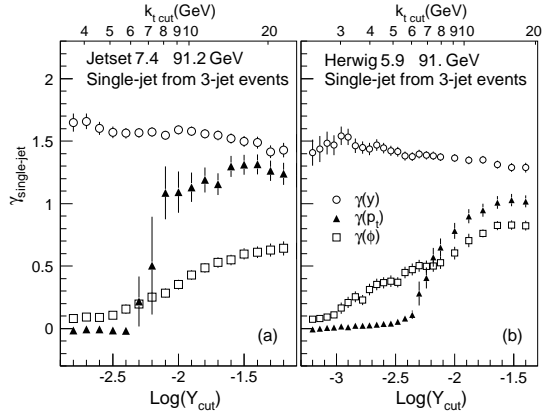}   
\caption{\label{Fig. 9} The variation of the parameters
$\gamma(y)$, $\gamma(\pt)$, $\gamma(\vf)$  with $\yct$ ($\ktct$)
in a single jet of 3-jet events  from (a) \js and (b) \hw at
$\sqrt s=$91.2 GeV, calculated in jet-axis frame.}
\end{figure}
\section{The dynamical fluctuations inside jets}

The transverse structure of jet has scaling property means that
the scale for jet production can be chosen within a wide range and
jets with a unique relative transverse momentum $\kt$ distribution
will still be obtained. This phenomenon supports the approach of
choosing jet production scale arbitrarily within a wide range.

Then how to understand the speciality of the {\it most reasonable}
scale of jet production at 4--6 GeV? This scale is related to the
QCD dynamics of jet production and development. Its special role
will be presented only in the dynamical fluctuations but not in
the particle distributions, \cf the discussion in Sec.\;III above.

In Fig.\;2 of Ref.\;\cite{CGLlsPRD} and Fig.\;2 of the present
paper the $\gamma$ parameters of event-dynamical-fluctuations have
been shown, where the factorial moments are calculated for the
{\it 2- or 3-jet events} in the {\it event-thrust frame}. Now we
want to study the dynamical fluctuations inside jets, so we take
the {\it particles in a single jet} and calculate the factorial
moments, \cf Appendix, of these particles in the {\it frame with
jet axis as the $z$ axis}. The results are shown in Fig's.\;8 and
9, where Fig's.\;8 is the variation of the parameters $\gamma(y)$,
$\gamma(\pt)$, $\gamma(\vf)$  with $\yct$ ($\ktct$) in a single
jet of 2-jet events and Fig's.\;9 is those in a single jet of
3-jet events from \js and \hw at $\sqrt s=$91.2 GeV, respectively.
All the figures are calculated in the jet-axis frame.

A transition point can clearly be seen from the figures, which is
located at $\ktct^{\rm dynam}= 6.84\pm 0.20$\;GeV (Jetset), $
5.62\pm 0.19$\;GeV (Durham) for single jet from 2-jet events; and
$\ktct^{\rm dynam}= 6.10\pm 0.19$\;GeV (Jetset),
 $ 5.82\pm 0.20$\;GeV (Durham) for that from 3-jet events.
 At this point the $\gamma(\vf)$ has a sudden
rise. This sudden rise from $\gamma(\vf)\sim 0$ to a saturation
value means that the gluons emitted with $\kt<\ktct^{\rm dynam}$
will be simply included in the parton shower, having no dynamical
fluctuation in azimuth. On the contrary, the gluons emitted with
$\kt>\ktct^{\rm dynam}$ will be {\it hard} enough to develop as a
separate jet and the dynamical fluctuations in azimuth appear.

\vskip5mm \centerline{Table~1 The scales for jet production}
\vskip2mm \noindent{\small\begin{tabular}{|c|c|c|}\hline
  &  \js  &  \hw \\
  \hline
$\ktct^{\rm circ}$ (GeV)  & $6.32\pm 0.03$ & $4.28\pm . 0.02$ \\
\hline
 $\ktct^{\mbox{\footnotesize 3-jet}}$ (GeV)
& $6.03\pm 0.06$ & $3.98\pm 0.03$  \\
\hline $\ktct^{\rm dynam}$  (GeV) from 2-jet events &$6.84\pm 0.20 $&$5.62\pm 0.19$ \\
\hline $\ktct^{\rm dynam}$  (GeV) from 3-jet events &$6.10\pm 0.19 $&$5.82\pm 0.20$ \\
\hline
\end{tabular}}
\vskip5mm

In Table\;I are listed the scales found in Ref.\;\cite{CGLlsPRD}, in
Sec.\;III and this section of the present paper.
The coincidence of
the scales obtained from different origins is remarkable. All of
them are related to jet production, so we give them a unique name
------ {\it jet production scale} and denote them as $\ktctprod$.

\section{Conclusion}

As discussed in Sec.\;III, a highly-virtual parton will emit
gluons and develop to a parton shower, hadronizing eventually to
jet. The jets are defined with a scale $\ktct$, which can take
values in a wide range, \cf Fig.\;7. When $\ktct$ takes a large
value, \eg $\ktct=12$\;GeV, the emitted gluon might in principle
have a $\kt$ value almost as large as this value, \ie only a
little bit smaller than 12\;GeV, and still be a ``daughter'' of
the mother parton. In other words, when $\ktct=12$\;GeV the
development of a parton to jet might be extended to a scale about
equal to 12\;GeV. However, it does not act as that. The
development of a parton to jet stops at a scale much smaller than
12\;GeV, \cf Fig's.\; 5, 6. We will refer to this scale as the
{\it jet development scale} and denote it as $\ktctdev$.

This scale measures the largest boundary of a parton shower
developed from a mother parton. Since the $\kt$ distribution tends
to zero exponentially, \cf Fig's.\;5, 6, we take the $\kt$ value,
where the probability density reduces for an order of magnitude
from its maximum, as a measure of $\ktctdev$, \ie we define
$\ktctdev$ through \beq \frac{1}{N} \frac{\d N}{\d \kt}
\f(\ktctdev\g) = 0.1 \cdot \f(\frac{1}{N} \frac{\d N}{\d
\kt}\g)_{\rm max}. \eeq Taking the 24 $\kt$ distributions obtained
for single jets in 2- and 3-jet events from two MC generators
--- \js and \hw, for 6 $\ktct$ values --- $\ktct=$ 2, 4, 6,
8, 10, 12 GeV, shown in Fig's.\;5, 6, as a unique sample and
calculate $\ktctdev$ from this sample according to Eq.\;(4), we
get \beq \ktctdev = 1.81 \pm 0.10\; \rm GeV. \eeq The highly
convergent result from events with different numbers of jet, from
differnt generators and from various values of $\ktct$ is
remarkable. It shows that the dynamics for the development of a
parton to jet is universal. It has a scale $\ktctdev$. All the
jets produced with various $\ktct$ is of the same scale.

However, the jets produced with different $\ktct$ are not totally
equivalent. There is a jet-production  scale $\ktctprod$, the jets
produced with $\ktct=\ktctprod$ are the most consistent with QCD
jet-production dynamics in the following sense:

\btm

\item The jets in 2-jet events produced with $\ktct=\ktctprod$ is
circular with respect to dynamical fluctuations, \ie the dynamical
fluctuations in these jets is circular in the  transverse plane.
These jets have axial dynamical symmetry as expected by QCD.

\item When a third jet is produced with $\ktct=\ktctprod$, the
resulting 3-jet system is isotropic in consistent with QCD
expectation.

\item Inside a single jet when a gluon is emitted with $\kt =
\ktctprod$, dynamical fluctuations in azimuthal angle appears
suddenly, which means that this gluon is no more a part of the
original jet but is the ``mother'' of a new jet.

\etm\vskip0mm

We conclude that there exist two scales in jet physics. One is the
{\it jet-development scale} $\ktctdev$ ($\sim$\;1.8\;GeV), which
determines the size of the jet developed from a mother-parton. The
other one is the {\it jet-production scale} $\ktctprod$
($\sim$\;4--6\;GeV), the jets produced with this scale are the
most consistent with QCD jet-production dynamics.

We can identify jets with various scales $\ktct$ in a wide range
and universal jets with the same transverse distribution will be
obtained. Among them only those identified with $\ktct=\ktctprod$
are the most consistent with QCD jet-production dynamics.  It is
the {\it best representative} of the mother parton. When we use
jets to study the properties of partons, those identified with
$\ktct=\ktctprod$ will provide the most reliable dynamical
information about their mother-partons.

\vskip8mm \akgt This work is supported by the Sci-tech Innovation
for Excellent University of Hubei No EJK0316 , the National
Science Foundation of China under project 10375025 and by the
Cultivation Fund of the Key Scientific and Technical Innovation
Project, Ministry of Education of China NO CFKSTIP-704035.

\appendix*

\section{Dynamical fluctuations}
The dynamical fluctuations can be characterized by the anomalous
scaling of normalized factorial moments (NFM)~\cite{BP}:
\bea   
  F_q(M)&=&{\frac {1}{M}}\sum\limits_{m=1}^{M}{\frac{\langle n_m(n_m-1)
     \cdots (n_m-q+1)\rangle }{{\langle n_m \rangle}^q}}
     \nonumber\\
   &\propto& (M)^{\phi_q}\ \  \quad \quad (M\to \infty) \ \ ,
\eea where a region $\Delta$ in 1-, 2- or 3-dimensional phase
space is divided into $M$ cells, $n_m$  is the multiplicity in the
$m$th cell, and $\langle\cdots\rangle$ denotes vertically
averaging over the event sample.

When the fluctuations exist in higher-dimensional (2-D or 3-D)
space, the projection effect~\cite{Ochs} will cause the
second-order 1-D NFM to go to saturation according to the rule
\cite{ftnt}:
\beq   
 F_2^{(a)}(M_a) = A_a-B_a M_a^{-\gamma_a}, \ \
\eeq where $a=1,2,3$ denotes the different 1-D variables. The
parameter $\gamma_a$ describes the rate of approach to saturation
of the NFM in direction $a$ and is the most important
characteristic for the higher-dimensional dynamical fluctuations.
If $\gamma_a = \gamma_b$, the fluctuations are isotropic in the
$a,b$ plane. If $\gamma_a \neq \gamma_b$, the fluctuations are
anisotropic in this plane.

\ed
\begin{thebibliography}{99}
\bibitem{natural} The natural unit system with $\hbar=c=1$ will be
used throughout in the present paper.
\bibitem{ftnt} In order to eliminate the influence of momentum
conservation~\cite{MMCN}, the first few points ($M=1,2$ or 3)
should be omitted when fitting the data to Eq.(3).
\end{thebibliography}
